\documentclass[prd,aps, tightenlines, preprintnumbers, showpacs, nofootinbib,superscriptaddress,notitlepage]{revtex4-1}
\pdfoutput=1

\usepackage{amssymb}
\usepackage{amsmath}
\usepackage[boldvectors,units]{physymb}

\usepackage{tikz}
\usepackage{pgfplots}
\usepackage{pgfplotstable}
\usetikzlibrary{calc}
\usetikzlibrary{decorations.pathmorphing}
\usetikzlibrary{decorations.markings}
\usetikzlibrary{external}
\usetikzlibrary{patterns}
\usetikzlibrary{shapes.geometric}
\usepgfplotslibrary{groupplots}

\tikzset{external/only named=true}
\tikzset{quark/.style={draw,postaction={decorate},decoration={markings,mark=at position .55 with {\arrow{>}}}}}
\tikzset{antiquark/.style={draw,postaction={decorate},decoration={markings,mark=at position .55 with {\arrow{<}}}}}
\tikzset{lepton/.style={draw,postaction={decorate},decoration={markings,mark=at position .55 with {\arrow{>}}}}}
\tikzset{antilepton/.style={draw,postaction={decorate},decoration={markings,mark=at position .55 with {\arrow{<}}}}}
\tikzset{gluon/.style={decorate,decoration={coil,amplitude=3pt,segment length=5pt}}}
\tikzset{small gluon/.style={decorate,decoration={coil,amplitude=1pt,segment length=2pt}}}
\tikzset{photon/.style={decorate,decoration={snake,amplitude=2pt,segment length=5pt}}}
\tikzset{weak/.style={decorate,decoration={zigzag,amplitude=3pt,segment length=5pt}}}

\newlength\baryonlinespacing
\tikzset{baryon/.style={to path={-- (\tikztotarget) \tikztonodes},
  execute at begin to={
   \draw ($(\tikztostart)!\baryonlinespacing!90:(\tikztotarget)$)--($(\tikztotarget)!\baryonlinespacing!-90:(\tikztostart)$);
   \draw[quark] (\tikztostart)--(\tikztotarget);
   \draw ($(\tikztostart)!\baryonlinespacing!-90:(\tikztotarget)$)--($(\tikztotarget)!\baryonlinespacing!90:(\tikztostart)$);
  }}
 }
\tikzset{right half baryon/.style={to path={-- (\tikztotarget) \tikztonodes},
  execute at begin to={
   \draw[quark] (\tikztostart)--(\tikztotarget);
   \draw ($(\tikztostart)!\baryonlinespacing!-90:(\tikztotarget)$)--($(\tikztotarget)!\baryonlinespacing!90:(\tikztostart)$);
  }}
 }
\tikzset{left half baryon/.style={to path={-- (\tikztotarget) \tikztonodes},
  execute at begin to={
   \draw ($(\tikztostart)!\baryonlinespacing!90:(\tikztotarget)$)--($(\tikztotarget)!\baryonlinespacing!-90:(\tikztostart)$);
   \draw[quark] (\tikztostart)--(\tikztotarget);
  }}
 }
\tikzset{nucleus/.style={to path={-- (\tikztotarget) \tikztonodes},
  execute at begin to={
   \draw ($(\tikztostart)!1.5\baryonlinespacing!90:(\tikztotarget)$)--($(\tikztotarget)!1.5\baryonlinespacing!-90:(\tikztostart)$);
   \draw ($(\tikztostart)!1.2\baryonlinespacing!90:(\tikztotarget)$)--($(\tikztotarget)!1.2\baryonlinespacing!-90:(\tikztostart)$);
   \draw ($(\tikztostart)!0.9\baryonlinespacing!90:(\tikztotarget)$)--($(\tikztotarget)!0.9\baryonlinespacing!-90:(\tikztostart)$);
   \draw ($(\tikztostart)!0.3\baryonlinespacing!90:(\tikztotarget)$)--($(\tikztotarget)!0.3\baryonlinespacing!-90:(\tikztostart)$);
   \draw[quark] (\tikztostart)--(\tikztotarget);
   \draw ($(\tikztostart)!0.3\baryonlinespacing!-90:(\tikztotarget)$)--($(\tikztotarget)!0.3\baryonlinespacing!90:(\tikztostart)$);
   \draw ($(\tikztostart)!0.9\baryonlinespacing!-90:(\tikztotarget)$)--($(\tikztotarget)!0.9\baryonlinespacing!90:(\tikztostart)$);
   \draw ($(\tikztostart)!1.2\baryonlinespacing!-90:(\tikztotarget)$)--($(\tikztotarget)!1.2\baryonlinespacing!90:(\tikztostart)$);
   \draw ($(\tikztostart)!1.5\baryonlinespacing!-90:(\tikztotarget)$)--($(\tikztotarget)!1.5\baryonlinespacing!90:(\tikztostart)$);
  }}
 }
\tikzset{blob/.style={ellipse,minimum height=0.8cm,minimum width=0.3cm}}
\tikzset{interaction/.style={ellipse,minimum height=0.3cm,minimum width=0.3cm}}

\pgfplotsset{compat=newest}

\pgfplotsset{
 use lhcf plot range/.style={
  ymax=1.3,
  ymin=1e-4,
  xmin=0,
  xmax=0.6
 },
 result axis/.style={
  ymin=1e-6,
  xlabel={$p_\perp [\si{GeV}]$},
  ylabel={$\frac{\udddc N}{\udc \eta\uddc p_\perp} \bigl[\si{GeV^{-2}}\bigr]$},
  legend style={cells={anchor=west}},
  legend columns=1,
  y filter/.code={\ifx\pgfmathresult\empty\def\pgfmathresult{-40}\fi}
 },
 data plot/.style={
  black,line width={0.05pt},mark=*,mark size={0.9pt},error bars/y dir=both,error bars/y explicit
 }
}

\pgfplotsset{
 LO/.style={
  /pgfplots/result plot fill/.style={
   /tikz/draw=none,
   /tikz/preaction={
    /tikz/fill=#1,
    /tikz/fill opacity=0.2
   },
   /tikz/pattern=crosshatch,
   /tikz/pattern color=#1,
   /pgfplots/area legend
  },
  /pgfplots/result plot line/.style={
   /tikz/draw=#1,
   /tikz/mark=x,
   /tikz/mark size={0.5pt},
   /tikz/mark options={fill=#1},
   /pgfplots/forget plot
  }
 },
 NLO/.style={
  /pgfplots/result plot fill/.style={
   /tikz/draw=none,
   /tikz/fill=#1,
   /tikz/fill opacity=0.6,
   /pgfplots/area legend
  },
  /pgfplots/result plot line/.style={
   /tikz/draw=#1,
   /tikz/mark=x,
   /tikz/mark size={0.5pt},
   /tikz/mark options={fill=#1},
   /pgfplots/forget plot
  }
 },
 exact/.style={
  /pgfplots/result plot fill/.style={
   /tikz/draw=none,
   /tikz/preaction={
    /tikz/fill=#1,
    /tikz/fill opacity=0.2
   },
   /tikz/pattern=crosshatch dots,
   /tikz/pattern color=#1,
   /pgfplots/area legend
  },
  /pgfplots/result plot line/.style={
   /tikz/draw=#1,
   /tikz/mark=x,
   /tikz/mark size={0.5pt},
   /tikz/mark options={fill=#1},
   /pgfplots/forget plot
  }
 },
 rcBK GBW LO/.style={
  LO=cyan!50!blue
 },
 rcBK GBW NLO/.style={
  NLO=red
 },
 rcBK GBW exact/.style={
  exact=green!50!black!70!purple
 },
}

\newcommand\fixedcoupling[1]{1}
\newcommand\runningcoupling[1]{pi/(2.25*(ln(greater(#1,1)*#1+notgreater(#1,1)*1) - ln(0.0588)))/0.2}
\newcommand\coupling[1]{\runningcoupling{#1}}

\newcommand\resultplot[4]{
 { % begin a TeX scope
  \expandafter\pgfplotstablevertcat\expandafter\filleddata\csname #4filled\endcsname
  \expandafter\pgfplotstablevertcat\expandafter\himudata\csname #4himu\endcsname
  \expandafter\pgfplotstablevertcat\expandafter\lomudata\csname #4lomu\endcsname
  \addplot[#1,result plot fill] table[x expr=#2,y expr=#3] {\filleddata};
 \addplot[#1,result plot line] table[x expr=#2,y expr=#3] {\himudata};
 \addplot[#1,result plot line,forget plot] table[x expr=#2,y expr=#3] {\lomudata};
 }
}

\def\readdata#1#2#3#4{
 \def\filenamepattern##1{#1}
 \pgfplotsset{table/sort cmp={float <}}
 \expandafter\pgfplotstablesort\csname #4himu\endcsname{\filenamepattern{#3}}
 \pgfplotsset{table/sort cmp={float >}}
 \expandafter\pgfplotstablesort\csname #4lomu\endcsname{\filenamepattern{#2}}
 \expandafter\pgfplotstablevertcat\csname #4filled\expandafter\endcsname\csname #4himu\endcsname
 \expandafter\pgfplotstablevertcat\csname #4filled\expandafter\endcsname\csname #4lomu\endcsname
}

\readdata{anc/brahms-dAu-mu#1-rcBKL01.Y22.summary.dat}{10}{50}{brahmsrcBKYA}
\readdata{anc/brahms-dAu-mu#1-rcBKL01.Y32.summary.dat}{10}{50}{brahmsrcBKYB}
\readdata{anc/brahms-dAu-mu#1-rcBKL01-e.Y22.summary.dat}{10}{50}{brahmsrcBKYC}
\readdata{anc/brahms-dAu-mu#1-rcBKL01-e.Y32.summary.dat}{10}{50}{brahmsrcBKYD}
\readdata{anc/star-dAu-mu#1-rcBKL01.Y4.summary.dat}{10}{50}{starrcBK}
\readdata{anc/star-dAu-mu#1-rcBKL01-e.Y4.summary.dat}{10}{50}{starrcBKe}

\pgfplotstablesort[row sep=\\]\brahmsdAuloY{
pt          ptlo                pthi                yield       staterr         syserr         \\
0.65        0.6                 0.7                 0.388093    0.00243152      0.0582139      \\
0.75        0.7                 0.8                 0.241047    0.00225041      0.036157       \\
0.85        0.8                 0.9                 0.165855    0.000938941     0.0248783      \\
0.95        0.9                 1                   0.112226    0.00107049      0.0168339      \\
1.09        1                   1.2                 0.0611503   0.000486684     0.00917254     \\
1.29        1.2                 1.4                 0.0299976   0.000307453     0.00449965     \\
1.49        1.4                 1.6                 0.0136606   0.000190933     0.00204909     \\
1.69        1.6                 1.8                 0.0068023   0.000125613     0.00102034     \\
1.89        1.8                 2                   0.00399323  9.32567e-05     0.000598985    \\
2.13        2.0                 2.3                 0.00205668  5.16416e-05     0.000308502    \\
2.55        2.3                 2.9                 0.000565847 9.04178e-06     8.48771e-05    \\
3.33        2.9                 4.0                 9.39018e-05 6.84736e-06     1.40853e-05    \\
}
\pgfplotstablesort[row sep=\\]\brahmsdAuhiY{
pt          ptlo                pthi                yield       staterr         syserr         \\
0.25        0.2                 0.3                 2.01147     0.0258206       0.3017205      \\
0.35        0.3                 0.4                 1.45906     0.0174999       0.218859       \\
0.45        0.4                 0.5                 0.933742    0.0114546       0.1400613      \\
0.55        0.5                 0.6                 0.466025    0.00760817      0.06990375     \\
0.65        0.6                 0.7                 0.293151    0.00555747      0.04397265     \\
0.75        0.7                 0.8                 0.142542    0.00358952      0.0213813      \\
0.85        0.8                 0.9                 0.085915    0.00174568      0.01288725     \\
0.95        0.9                 1                   0.0559974   0.00123521      0.00839961     \\
1.09        1                   1.2                 0.0233523   0.000140574     0.00467046     \\
1.29        1.2                 1.4                 0.0104938   8.29051e-05     0.00209876     \\
1.49        1.4                 1.6                 0.00506653  4.97884e-05     0.001013306    \\
1.69        1.6                 1.8                 0.00246804  3.15367e-05     0.000493608    \\
1.89        1.8                 2                   0.00120714  1.93553e-05     0.000241428    \\
2.13        2                   2.3                 0.000459623 9.66696e-06     6.89434e-05    \\
2.55        2.3                 2.9                 0.000124681 3.42328e-06     1.870215e-05   \\
3.31        2.9                 4                   1.22812e-05 7.80291e-07     1.84218e-06    \\
}
\newcommand\stardAusigmainel{2.21e6} % inelastic cross section for dAu @ STAR: 2.21b
\pgfplotstablesort[row sep=\\]\stardAu{
 epilo          epihi           epi         pt          xsec        staterr     syserr \\
 25             30              27.1        1.03        9900        100         1500   \\
 30             35              32.2        1.21        2750        40          290    \\
 35             40              37.2        1.38        970         20          120    \\
 40             45              42.1        1.55        315         11          48     \\
 45             50              47.1        1.72        110         6           25     \\
 50             55              52.0        1.91        46          4           10     \\
}

\newcommand\hadronconversionfactor{1.3}

\newcommand\pp{\ensuremath{\mathrm{pp}}}
\newcommand\pA{\ensuremath{\mathrm{p}A}}
\newcommand\dA{\ensuremath{\mathrm{d}A}}
\newcommand\dAu{\ensuremath{\mathrm{dAu}}}

\begin{document}
\title{Matching Collinear and Small-$x$ Factorization Calculations for Inclusive Hadron Production in $\pA$ Collisions}

\author{Anna M. Sta\'sto}
\affiliation{Department of Physics, Pennsylvania State University, University Park, PA 16802, USA}
\affiliation{Institute of Nuclear Physics, Polish Academy of Sciences, ul. Radzikowskiego 152, Krak\'ow, Poland}
\author{Bo-Wen Xiao}
\affiliation{Key Laboratory of Quark and Lepton Physics (MOE) and Institute
of Particle Physics, Central China Normal University, Wuhan 430079, China}

\author{Feng Yuan}
\affiliation{Nuclear Science Division, Lawrence Berkeley National
Laboratory, Berkeley, CA 94720, USA}

\author{David Zaslavsky}
\affiliation{Department of Physics, Pennsylvania State University, University Park, PA
16802, USA}

\begin{abstract}
We construct a theoretical framework to match the formulas for forward inclusive hadron productions in $\pA$ collisions in the small-$x$ saturation formalism and collinear factorization.
The small-$x$ calculation can be viewed as a power series in $Q_s^2/k_\perp^2$, in which the collinear factorization result corresponds to the leading term.
At high transverse momentum, the subleading correction terms are insignficant, whereas at low $p_\perp$, the power corrections become important and the small-$x$ resummation is essential to describe the differential cross section.
We show that the familiar collinear factorization calculation can smoothly match the results from small-$x$ factorization at next-to-leading order in $\alpha_s$ when we use exact kinematics, as opposed to the approximate kinematics in previous work.
With this matching, we can describe the experimental data from RHIC very well at high $p_\perp$.
\end{abstract}
\maketitle

\section{Introduction}

Single inclusive hadron production in $\pA$ (as opposed to $\pp$) collisions has long been regarded as one of the best signals for strong nuclear effects in high energy scattering, in particular for gluon saturation at small $x$ in a large nucleus.
Recently, collider and detector technology has advanced enough to make the small-$x$ kinematic region accessible, and accordingly within the past decade, there has been a concerted effort to detect single inclusive production. Studies at RHIC~\cite{Arsene:2004ux, Adams:2006uz, Braidot:2010ig, Adare:2011sc} and the LHC~\cite{ALICE:2012xs, ALICE:2012mj, Hadjidakis:2011zz, LHCb} have already provided plenty of data at middle and forward rapidities.

On the theoretical side, there has been a concurrent effort to understand and predict the results from the RHIC and LHC experiments. Gluon saturation has been argued
to play a very important role in describing the experimental data from RHIC, which display suppression in forward $\dA$ collisions~\cite{Dumitru:2002qt,Albacete:2013ei, Dumitru:2005kb, Albacete:2010bs, Levin:2010dw, Fujii:2011fh, Albacete:2012xq, Lappi:2013zma}.
The theoretical calculation of this cross section has been carried out in the small-$x$ factorization formalism up to next-to-leading order (NLO)~\cite{Chirilli:2011km, Chirilli:2012jd}.

However, it is nontrivially difficult to compute the single inclusive hadron productions over a wide range of transverse momenta $p_\perp$ in a single framework.
This is because the small-$x$ formalism at NLO, which works for the low $p_\perp$ region, yields a negative cross section\footnote{
The negative cross section found at high $p_\perp$ is not too surprising since it is well-known that fixed order calculations beyond leading order are not guaranteed to be positive.
Although this is not a theoretical problem, it is indeed troublesome for small-$x$ physics in phenomenology.}
in the large $p_\perp$ regime~\cite{Stasto:2013cha}, while the conventional collinear factorization QCD calculation, which can describe the high $p_\perp$ region, often overpredicts the low $p_\perp$ data, as our results will show.
We can attribute this to different physics acting in the two regions.
At low $p_\perp$, the longitudinal momentum fraction of partons actively involved in the scattering is relatively small, and the parton densities in nuclei are relatively high;
therefore, the transverse momentum of the produced hadron comes from the accumulated effect of multiple scatterings from gluons in the target nucleus, which in turn are influenced by the small-$x$ evolution of the parton density (the Balitsky-Kovchegov or Jalilian-Marian-Iancu-McLerran-Weigert-Leonidov-Kovner evolution~\cite{Balitsky:1995ub, Kovchegov:1999yj, Jalilian-Marian:1997jx+X}).
On the other hand, when $p_\perp$ is large, hard scattering becomes the main contribution to the transverse momentum in the final state, and multiple scatterings are negligible; therefore the collinear QCD calculation should  be better justified in that region.

Clearly, the predictive power of both models would be improved by systematically matching the small-$x$ calculation in the low-$p_\perp$ region to the collinear factorization calculation in the high-$p_\perp$ region. 
This issue has been partially studied in Refs.~\cite{Dumitru:2002ha, Altinoluk:2011qy}.
That line of research has led to the conclusion that hard gluon radiation will dominate the differential cross section for high-$p_\perp$ particle production in $pA$ collisions.
Previous studies compensated for this effect by assuming that the dipole gluon distribution, used in the small-$x$ factorization approach, has a perturbative power tail.
This is sufficient to bring power-like behavior for the $p_\perp$ spectrum, but it does not enable the small-$x$ formalism to describe inclusive hadron production at large transverse momenta.
A simple modification of the perturbative tail in the dipole gluon distribution will undoubtedly introduce theoretical uncertainties in the predictions.
That is indeed what happened in the phenomenological small-$x$ calculations for these observables in the last few years prior to $\pA$ collisions at the LHC.

The goal of this paper is to build a consistent and rigorous framework to match the small-$x$ saturation formalism and the collinear factorization for forward inclusive hadron productions in $\pA$ collisions.
We find that by taking the exact kinematics for finite $\sqrt{s}$ into account, we obtain a result from collinear factorization which smoothly matches to the next-to-leading order small-$x$ factorization result. 
Under the exact kinematics, we find complete agreement for all partonic channels between these two formalisms at sufficiently high energy when t-channel gluon exchanges become dominant. 

Our approach is based on theoretical calculations for two-particle production in forward $\pA$ collisions in the small-$x$ factorization formalism~\cite{Dominguez:2010xd, Dominguez:2011wm}.
It has been shown that they lead to a consistent picture for the differential cross sections in the so-called correlation limits, as compared to the collinear (and transverse momentum-dependent) factorization calculations for the same observables.
When we integrate out the phase space of one particle in the two particle differential cross section, we obtain the formula for single inclusive hadron production at large transverse momentum.
This naturally provides us a matching with the collinear factorization calculation for inclusive hadron production in $\pA$ collisions.
In particular, the collinear factorization result is the leading power expansion in $Q_s^2/k_\perp^2$ of the formulae derived in the small-$x$ formalism.
Through a detailed analytical comparison, we build a systematic and complete connection to the collinear factorization of the calculation in the small-$x$ formalism. This connection  will strengthen the predictive power of the calculations which take into account small-$x$ physics.

\section{Matching collinear factorization and small-$x$ factorization}

We start with the two-particle cross section, derived in Refs.~\cite{Chirilli:2011km, Chirilli:2012jd, Marquet:2007vb, Dominguez:2010xd, Dominguez:2011wm}, which exhibits perfect matching between the small-$x$ and collinear factorization results for the two final state particles at large transverse momenta. Integrating over the phase space of one of the particles gives us the single inclusive cross section at forward rapidity, with $y$ and $p_\perp$ defined as the rapidity and transverse momentum, respectively, of the produced hadron. 
We will then demonstrate that this matches the equivalent result from collinear factorization.

\subsection{Single inclusive production in the small-$x$ formalism}

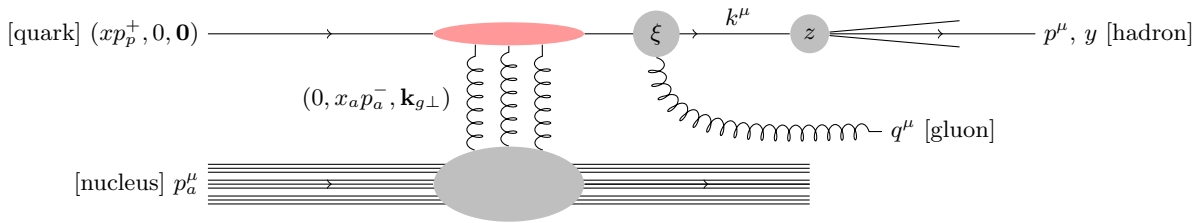
\begin{figure}
 \centering
 \tikzsetnextfilename{figure-diagram}
 \begin{tikzpicture}
  \node[ellipse,fill=red!40,minimum width=2cm] (interaction) at (0,0) {};
  \draw[quark] (interaction) +(-4,0) -- (interaction) node[pos=0,left] {[quark] $(x p_p^+, 0, \vec{0})$};
  \path (interaction) +(0,-2) node[ellipse,minimum width=2cm,minimum height=1cm] (gluon creation) {};
  \draw[gluon] (gluon creation.135) -- (interaction.198) node[pos=0.5,left=2mm] {$(0, x_a p_a^-, \vec{k}_{g\perp})$};
  \draw[gluon] (gluon creation.90) -- (interaction.270);
  \draw[gluon] (gluon creation.45) -- (interaction.342);
  \path (interaction) +(4,0) node[circle,minimum width=5mm,fill=gray!50] (fragmentation) {$z$};
  \draw[quark] (interaction) -- (fragmentation) node[pos=0.75,above] {$k^\mu$} node[pos=0.35,circle,minimum width=2mm,fill=gray!50] (gluon emission) {$\xi$};
  \draw[quark] (fragmentation) -- +(3,0) node[right] {$p^\mu$, $y$ [hadron]};
  \foreach \th in {-5,5} \draw (fragmentation) -- +(\th:2);
  \draw[nucleus] (gluon creation) +(-4,0) node[left] {[nucleus] $p_a^\mu$} to (gluon creation);
  \draw[nucleus] (gluon creation) to +(4,0);
  \draw[gluon] (gluon emission) to[out=270,in=180] +(3,-1.3) node[right] {$q^\mu$ [gluon]};
  \node[ellipse,minimum width=2cm,minimum height=1cm,fill=gray!50] at (gluon creation) {};
 \end{tikzpicture}
 \caption{A cartoon of the $2\to 2$ subprocess of the NLO $q\to qg$ channel interaction, with selected momenta and momentum fractions labeled, as well as the rapidity $y$ of the detected hadron. The momentum fractions $z$ and $\xi$ are defined by $p_\perp = zk_\perp$ and $k^+ = (1 - \xi) x p_p^+$, and $x$ is the longitudinal momentum fraction of the incoming quark with respect to the proton projectile.}
 \label{fig:qqgdiagram}
\end{figure}

For example, consider the $2\to 2$ subprocess in the $q\to qg$ channel with the final state quark fragmenting into the hadron, $q\to h(y, p_\perp)$, as shown in Figure~\ref{fig:qqgdiagram}. We integrate its cross section over the transverse momentum of the final state gluon by applying the delta function reflecting momentum conservation in the $2\to 2$ subprocess, which results in the differential cross section for single inclusive hadron production: % TODO: MUST make sure kinematic variables are consistent, in particular which momentum refers to which gluon?
\begin{equation}
\frac{\alpha _{s}}{2\pi ^{2}}\int \frac{\udc z}{z^{2}}D_{h/q}(z)\int_{\tau
/z}^{1}\udc\xi \frac{1+\xi ^{2}}{1-\xi }xq(x)\left\{ C_{F}\int \uddc\vec{k}_{g\perp }
\mathcal{I}(\vec{k}_\perp,\vec{k}_{g\perp})
+\frac{N_{c}}{2}\int \uddc \vec{k}_{g\perp }\uddc{\vec{k}_{g1\perp }}\mathcal{J}(\vec{k}_\perp,\vec{k}_{g\perp},\vec{k}_{g1\perp})\right\}\label{real} \; .
\end{equation}
The meanings of the variables $x$, $\tau$, and $\xi$ in this expression were given in Figure~\ref{fig:qqgdiagram}; they satisfy the kinematic relation $x=\tau /z\xi$, where $\tau=\frac{p_\perp e^{y}}{\sqrt{s}}$. We also have $C_F=(N_c^2-1)/2N_c$, and $\mathcal{I}$ and $\mathcal{J}$ are defined as
\begin{align}
\mathcal{I}(\vec{k}_\perp,\vec{k}_{g\perp})
&= \mathcal{F}_{x_a}(k_{g\perp}) \biggl[\frac{\vec{k}_\perp-\vec{k}_{g\perp}}{( \vec{k}_{\perp }-\vec{k}_{g\perp})^{2}}
-\frac{\vec{k}_\perp-\xi\vec{k}_{g\perp}}{(\vec{k}_{\perp }-\xi \vec{k}_{g\perp})^{2}}\biggr]^2, \\
\mathcal{J}(\vec{k}_\perp,\vec{k}_{g\perp},\vec{k}_{g1\perp})&=\biggl[ \mathcal{F}_{x_a}
(k_{g\perp})\delta ^{(2) }\left( \vec{k}_{g1\perp }-\vec{k}_{g\perp}\right) -
\mathcal{G}_{x_a}(\vec{k}_{g\perp},\vec{k}_{g1\perp })\biggr]\frac{2(\vec{k}_{\perp }-\xi
\vec{k}_{g\perp})\cdot (\vec{k}_{\perp }-\vec{k}_{g1\perp })}{(\vec{k}_{\perp }-\xi
\vec{k}_{g\perp})^{2}(\vec{k}_{\perp }-\vec{k}_{g1\perp })^{2}} \\
\text{with} \quad \mathcal{G}(\vec{k}_\perp,\vec{l}_{\perp})&=\int\frac{\uddc \vec{x}_\perp
\uddc\vec{y}_\perp \uddc\vec{b}_\perp}{(2\pi)^4}e^{-i\vec{k}_\perp\cdot
(\vec{x}_\perp-\vec{b}_\perp)-i\vec{l}_{\perp}\cdot(\vec{b}_\perp-\vec{y}_\perp)}
S_{x_a}^{(4)}(\vec{x}_\perp,\vec{b}_\perp,\vec{y}_\perp) \; ,
\end{align}
and $S_{x_a}^{(4)}(\vec{x}_\perp,\vec{b}_\perp,\vec{y}_\perp)=\frac{1}{N_c^2}\langle \mathrm{Tr}[{U}%
(\vec{x}_\perp){U}^\dagger(\vec{b}_\perp)] \mathrm{Tr}[{U}(\vec{b}_\perp){U}%
^\dagger(\vec{y}_\perp)]\rangle_{x_a}$.
To compute the cross section in $\dAu$ collisions, we will use the same formula; we assume that we can just include the additional parton distributions from the neutron according to isospin symmetry.

At large transverse momentum $k^2_\perp\gg Q_s^2$, we can expand the differential cross section from Eq.~\eqref{real} in terms of $Q^2_s/k_\perp^2$, and find that it exactly reproduces the collinear factorization calculation.
As a first step, keeping the leading power contributions, we obtain 
\begin{align}
\frac{\udddc\sigma}{\udc y \uddc\vec{p}_\perp}&=\frac{\alpha _{s}}{2\pi ^{2}}\int \frac{\udc z}{z^{2}}D_{h/q}(z)\int_{\tau
/z}^{1}\udc\xi \frac{1+\xi ^{2}}{1-\xi }xq(x)\left\{ C_{F}\frac{(1-\xi)^2}{k_\perp^4}+N_c\frac{\xi}{k_\perp^4}\right\}
\int \uddc\vec{k}_{g\perp}k_{g\perp}^2\mathcal{F}_{x_a}(k_{g\perp})\\
&=\frac{\alpha _{s}^2}{N_c}\int \frac{\udc z}{z^{2}}D_{h/q}(z)\int_{\tau
/z}^{1}\udc\xi \frac{1+\xi ^{2}}{1-\xi }xq(x)\left\{ C_{F}\frac{(1-\xi)^2}{k_\perp^4}+N_c\frac{\xi}{k_\perp^4}\right\} 
x'G(x',\mu) \; ,
\label{expansion}
\end{align}
where $\vec{k}_\perp=\vec{p}_\perp/z$ as before, and the expressions for $x_a$ and $x'$ are given below in Eqs.~\eqref{exk}.
The dipole gluon distribution is normalized according to the following equations:
\begin{align}
\int \uddc\vec{k}_{g\perp} {\cal F}_{x_a}(k_{g\perp})&=S_\perp\ ,\\
\int \uddc\vec{k}_{g\perp} k_{g\perp}^2{\cal F}_{x_a}(k_{g\perp})&=S_\perp Q_s^2\simeq \frac{\alpha_s2\pi^2}{N_c}x^\prime G(x^\prime,\mu) \; ,
\label{intg}
\end{align}
where $S_\perp$ represents the transverse area of the target nucleus and $x^\prime G(x^\prime,\mu)$ is the integrated gluon distribution from the nucleus, with $\mu$ being the renormalization scale.
In the small-$x$ formalism, it is believed that $\mu\simeq Q_s$ as explicitly shown in Ref.~\cite{Chirilli:2011km, Chirilli:2012jd}. 

Before demonstrating the matching to the collinear factorization, let us first take a closer look at the kinematics in the small-$x$ formalism.
For $2\to 2$ processes, one can easily obtain the following exact kinematic relations from energy-momentum conservation (details are in appendix~\ref{sec:kinematicappendix}):
\begin{subequations}
\begin{align}
x&=\frac{k_\perp}{\sqrt{s}\xi}e^y\label{exk:x}\\
x_a&=\frac{k_\perp}{\sqrt{s}}e^{-y}+\frac{(\vec{k}_{g\perp}-\vec{k}_\perp)^2}{\sqrt{s} k_\perp}\frac{\xi}{1-\xi}e^{-y}\label{exk:xa}\\
x^\prime&=\frac{k_\perp}{\sqrt{s}}e^{-y}+\frac{k_\perp}{\sqrt{s}}\frac{\xi}{1-\xi}e^{-y},\label{exk:xprime}
\end{align}
\label{exk}
\end{subequations}
where the kinematic variables are to be interpreted as shown in Fig.~\ref{fig:qqgdiagram}.

Strictly speaking, the small-$x$ factorization derived in Ref.~\cite{Chirilli:2011km, Chirilli:2012jd} requires that the center-of-mass energy $s\to \infty$ while $x$ is kept large, which indicates that the forward rapidity $y$ should also be kept large to maintain the relation~\eqref{exk:x} between $x$, $\sqrt{s}$, and $y$.
In this limit, it is straightforward to see that $x_a\to 0$ as $s\to \infty$.
However, $\sqrt{s}$ is only $\SI{200}{GeV}$ at RHIC, which is not particularly large. 
In order to apply the small-$x$ calculation to phenomenology, therefore, we need to pay attention to the kinematics and ensure that the small-$x$ factorization is in fact applicable~\cite{Beuf:2014uia}.

In the analysis of the hybrid factorization in the small-$x$ formalism, Ref.~\cite{Chirilli:2011km, Chirilli:2012jd}, what we have done is take $\vec{k}_\perp$ to be in the vicinity of $\vec{k}_{g\perp}$, which is of the order of the saturation momentum $Q_s$ or less when the gluon distribution $\mathcal{F}(k_{g\perp})$ is saturated. Under this assumption, one can approximately write $x_a\simeq\hat{x}_a =\frac{k_\perp}{\sqrt{s}}e^{-y}$ because the second term in Eq.~\eqref{exk:xa} vanishes. However, when $k_\perp \gg Q_s$, as at high $p_\perp$, this is no longer valid. We need to keep the second term of Eq.~\eqref{exk:xa} in the expression, and as a result, the natural kinematic condition $x_a \leq 1$ puts a constraint on $\xi$:
\begin{equation}
 \xi \leq \frac{1 - \hat{x}_a}{1 - \hat{x}_a + \hat{x}_a(\vec{k}_{g\perp} - \vec{k}_\perp)^2/k_\perp^2}.
\end{equation}
This keeps $\xi$ less than $1$, except at the single point $\vec{k}_\perp = \vec{k}_{g\perp}$ which makes a negligible contribution to the integral. In the limit $k_\perp \gg k_{g\perp}$, the above constraint becomes $\xi\leq 1-\hat x_a$.
Here we are adopting a kinematical constraint on the hard factors at NLO, which is similar to the kinematical constraint discussed in Ref.~\cite{Beuf:2014uia} for small-$x$ evolutions.

In practice, we implement this constraint by limiting the integrals over $\vec{k}_{g\perp}$ in Eqs.~\eqref{expansion} and~\eqref{intg}\footnote{One might naively object that the integral in Eq.~\eqref{intg} should cover all of $\mathbb{R}^2$, but the physical processes contributing to the integrated gluon distribution $G$ must be limited to those which satisfy energy-momentum conservation, or equivalently, those in which the gluon momentum satisfies Eq.~\eqref{eq:kgperpcondition}.
So in exact kinematics, the region of integration in Eq.~\eqref{intg} must be the region in which Eq.~\eqref{eq:kgperpcondition} is fulfilled, namely $\mathcal{R}$.} to the region $\mathcal{R}$, defined as the set of all $\vec{k}_{g\perp}\in\realset^2$ which satisfy
\begin{equation}
 (\vec{k}_{g\perp} - \vec{k}_\perp)^2 \leq k_\perp\bigl(\sqrt{s} e^{y} - k_\perp\bigr)\frac{1 - \xi}{\xi}\label{eq:kgperpcondition}
\end{equation}
which is a circle in $\vec{k}_{g\perp}$ space centered on $\vec{k}_\perp$.
This condition follows directly from, and is equivalent to, $x_a \leq 1$, using the definition of $x_a$ in Eq.~\eqref{exk:xa}.
Accordingly, the integrals cover all kinematically allowed values of $k_{g\perp}$.

\subsection{Connection to collinear factorization}

The above identifications are key to connecting the small-$x$ calculations with those in the collinear factorization calculations.
In the collinear factorization calculations, we have the differential cross section from the $2\to 2$ subprocess contribution as
\begin{equation}
\frac{\udddc\sigma}{\udc y\uddc\vec{p}_\perp}=\int \frac{\udc z}{z^{2}}D_{h/q}(z)\int \udc x \udc x' q(x)G(x')\frac{|\overline{\cal M}|^2}{2\hat s}
\frac{1}{2(2\pi)^3}2\pi \delta\bigl(\hat s+\hat t+ \hat u\bigr)\ ,%\delta(xx'\hat s+x \hat t+x' \hat u) \ ,
\end{equation}
where the scattering amplitude squared is
\begin{equation}
|\overline{\cal M}|^2=\frac{g^4}{N_c}\left[-C_F\frac{\hat u^2+\hat s^2}{\hat s\hat u}+N_c\frac{\hat u^2+\hat s^2}{\hat t^2}\right] \ .
\end{equation}
To compare to the small-$x$ calculation, we will apply the following kinematic
identities:
\begin{equation}
\hat s=\frac{k_\perp^2}{\xi(1-\xi)},~~\hat t=-\frac{k_\perp^2}{\xi},~~\hat u=-\frac{k_\perp^2}{1-\xi} \ .
\end{equation}
Substituting the above equations, we will obtain the differential cross section contribution as
\begin{equation}
\frac{\udddc\sigma}{\udc y\uddc\vec{p}_\perp}=\frac{\alpha_s^2}{N_c} \int \frac{\udc z}{z^{2}}D_{h/q}(z)\int \frac{\udc \xi}{\xi}
xq(x)x'G(x')\frac{1+\xi^2}{1-\xi}\frac{\xi}{k_\perp^4}\left[C_F(1-\xi)^2+N_c\xi\right]\label{22}
\end{equation}
where $x'$ is defined in Eq.~(\ref{exk:xprime}).
This is exactly the same as Eq.~(\ref{expansion}), the leading power expansion of the small-$x$ calculation at large transverse momentum, for given $\mu$ in the large $N_c$ limit.
Evidently, keeping the complete kinematics as shown in Eqs.~(\ref{exk}) and thereby restricting $x_a$ to be less than $1$ allows us to match the small-$x$ factorization result to the collinear factorization calculations.

\subsection{Numerical results from combined channels}

We have now established that at large $k_\perp$, the following formula matches the small-$x$ factorization calculation to the collinear factorization calculation:
\begin{equation}
\frac{\udc\sigma_{q\to q}}{\udc y\uddc\vec{p}_\perp}=\frac{\alpha _{s}}{2\pi ^{2}}\int^1_{\tau} \frac{\udc z}{z^{2}}D_{h/q}(z)\int_{\tau/z}^{1}\udc\xi \frac{1+\xi ^{2}}{1-\xi }xq(x)\left\{ C_{F}\frac{(1-\xi)^2}{k_\perp^4}+N_c\frac{\xi}{k_\perp^4}\right\}
\int_{\mathcal{R}} \uddc\vec{k}_{g\perp }k_{g\perp}^2\mathcal{F}_{x_a}(k_{g\perp}), \label{q1}
\end{equation}
We can now use this formula to compute the high-$p_\perp$ spectrum for the $q\to q$ channel.

Similarly, for the $g\to g$ channel, we can compute accordingly, in the large $N_c$ limit, 
\begin{equation}
\frac{\udddc\sigma_{g\to g}}{\udc y\uddc\vec{p}_\perp}=\frac{\alpha_s N_c }{2\pi^2}\int^1_{\tau} \frac{\udc z}{z^2} D_{h/g}(z)\int_{\tau/z}^{1}\udc\xi x g(x) \frac{2 [1-\xi (1-\xi)]^2 [1+\xi^2+(1-\xi)^2]}{ \xi (1-\xi)}
\frac{1}{k_\perp^4}\int_{\mathcal{R}} \uddc \vec{k}_{g\perp }k_{g\perp}^2\mathcal{F}_{x_a}(k_{g\perp}).
\end{equation}
To build a complete and systematic connection between the collinear factorization and saturation formalism at high transverse momentum limit, let us continue to examine all the possible  channels included in the collinear factorization and comment on their corresponding contributions in the saturation formalism. 

At sufficiently high energy, it is well-known that the $t$-channel gluon exchange graphs dominate the cross section~\cite{Kovchegov:2012mbw}, which allows us to neglect the  channels in which $q$ or $\bar q$ are exchanged, such as the $q\bar q \to gg$ and $gg\to q\bar q$ channels.
With the t-channel gluon exchange in mind, one should compute the $q q^\prime \to qq^\prime$, $gq^\prime\to gq^\prime$, $qg^\prime\to qg^\prime$ and $gg^\prime\to gg^\prime$ channels in the collinear factorization.
Here we use $q,g$ to denote incoming partons from the projectile proton, while $q^\prime, g^\prime$ represent partons from the target nucleus.
As demonstrated above, after using the same kinematics, we can find agreement between the collinear factorization result for $qg^\prime\to qg^\prime$ and the NLO saturation result for $q\to q$, and also between the collinear $gg^\prime\to gg^\prime$ result and the $g\to g$ channel in the saturation formalism. Furthermore, we find that the $qq^\prime \to qq^\prime$ and $gq^\prime\to gq^\prime$ channels in the collinear factorization correspond to the leading order $q\to q$ and $g\to g$ contributions in the small-$x$ formalism, respectively.
Finally, the off-diagonal channels $q\to g$ and $g\to q(\bar q)$ in the saturation formalism, which are always positive and numerically negligible, are related to the $q\bar q \to gg$ and $gg\to q\bar q$ channels in the collinear factorization. 

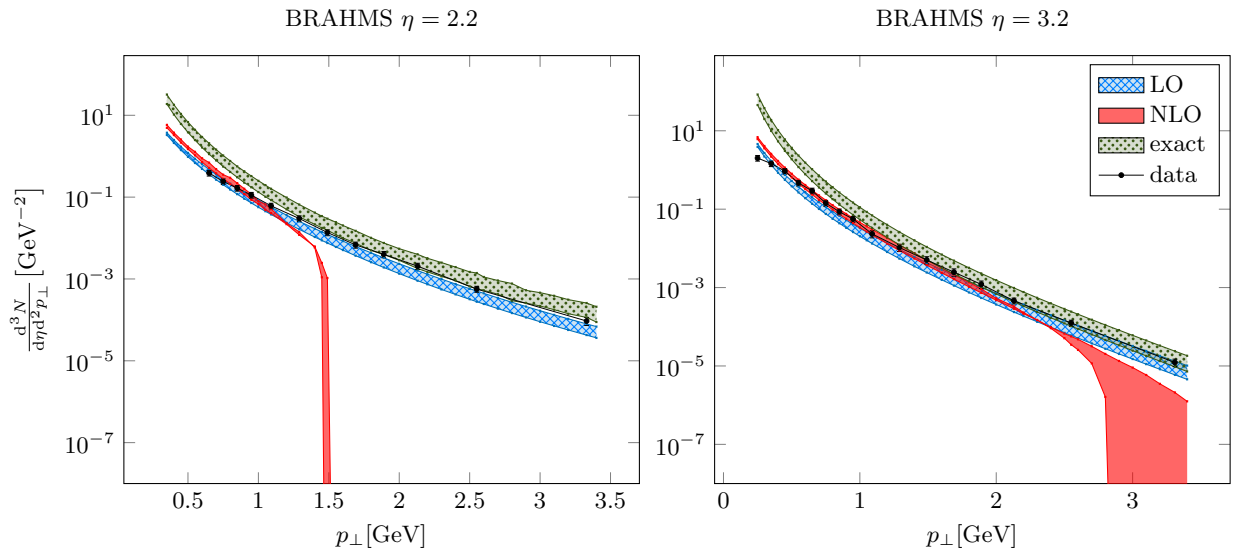
\begin{figure}[tbp]
 \centering
 \tikzsetnextfilename{figure-brahms}
 \begin{tikzpicture}
  \begin{groupplot}[result axis,ymin=1e-8,ymode=log,group style={group size=2 by 1,ylabels at=edge left}]
  \nextgroupplot[title={BRAHMS $\eta=2.2$}]
   \resultplot{rcBK GBW LO}{\thisrow{pT}}{\hadronconversionfactor*\thisrow{lomean}}{brahmsrcBKYA}
   \resultplot{rcBK GBW NLO}{\thisrow{pT}}{\hadronconversionfactor*(\thisrow{lomean}+\coupling{\thisrow{mu2}}*\thisrow{nlomean})}{brahmsrcBKYA}
   \resultplot{rcBK GBW exact}{\thisrow{pT}}{\hadronconversionfactor*(\thisrow{lo-mean}+\coupling{\thisrow{mu2}}*\thisrow{nlo-mean})}{brahmsrcBKYC}
   \addplot[data plot] table[x=pt,y expr={\thisrow{yield}},y error expr={(\thisrow{staterr} + \thisrow{syserr})}] {\brahmsdAuloY};
  \nextgroupplot[title={BRAHMS $\eta=3.2$}]
   \resultplot{rcBK GBW LO}{\thisrow{pT}}{\hadronconversionfactor*\thisrow{lomean}}{brahmsrcBKYB}
   \resultplot{rcBK GBW NLO}{\thisrow{pT}}{\hadronconversionfactor*(\thisrow{lomean}+\coupling{\thisrow{mu2}}*\thisrow{nlomean})}{brahmsrcBKYB}
   \resultplot{rcBK GBW exact}{\thisrow{pT}}{\hadronconversionfactor*(\thisrow{lo-mean}+\coupling{\thisrow{mu2}}*\thisrow{nlo-mean})}{brahmsrcBKYD}
   \addplot[data plot] table[x=pt,y expr={\thisrow{yield}},y error expr={(\thisrow{staterr} + \thisrow{syserr})}] {\brahmsdAuhiY};
   \legend{LO,NLO,exact,data}
  \end{groupplot}
 \end{tikzpicture}
\caption[*]{Comparison between the BRAHMS $h^-$ data~\cite{Arsene:2004ux} at pseudorapidities $\eta = 2.2,3.2$ and the LO and NLO small-$x$ computations~\cite{Stasto:2013cha, SOLO}, as well as the large $p_\perp$ perturbative results with exact kinematics, at $y = 2.2,3.2$. The edges of the band were computed with $\mu^2 = \SI{10}{GeV^2}$ and $\mu^2 = \SI{50}{GeV^2}$, thus the width of the band indicates the theoretical uncertainty due to the factorization scale. Calculated results use the rcBK gluon distribution.}
\label{ex}
\end{figure}

\begin{figure}[tbp]
 \centering
 \tikzsetnextfilename{figure-star}
 \begin{tikzpicture}
  \begin{axis}[result axis,ymin=1e-6,ymode=log,title={STAR $\eta=4$}]
   \resultplot{rcBK GBW LO}{\thisrow{pT}}{\thisrow{lomean}}{starrcBK}
   \resultplot{rcBK GBW NLO}{\thisrow{pT}}{\thisrow{lomean}+\coupling{\thisrow{mu2}}*\thisrow{nlomean}}{starrcBK}
   \resultplot{rcBK GBW exact}{\thisrow{pT}}{\thisrow{lo-mean}+\coupling{\thisrow{mu2}}*(\thisrow{nlo-mean})}{starrcBKe}
   \addplot[data plot] table[x=pt,y expr={\thisrow{xsec}/\stardAusigmainel},y error expr={(\thisrow{staterr} + \thisrow{syserr})/\stardAusigmainel}] {\stardAu};
   \legend{LO,NLO,exact,data}
  \end{axis}
 \end{tikzpicture}
\caption[*]{Comparison between the STAR $\pi^0$ data~\cite{Adams:2006uz} at pseudorapidity $\eta=4$ and the LO and NLO small-$x$ calculations~\cite{Stasto:2013cha, SOLO}, as well as the large $p_\perp$ perturbative results with exact kinematics, at $y=4$. As in Fig.~\ref{ex}, the edges of the bands show $\mu^2 = \SI{10}{GeV^2}$ and $\SI{50}{GeV^2}$.}
\label{ex2}
\end{figure}
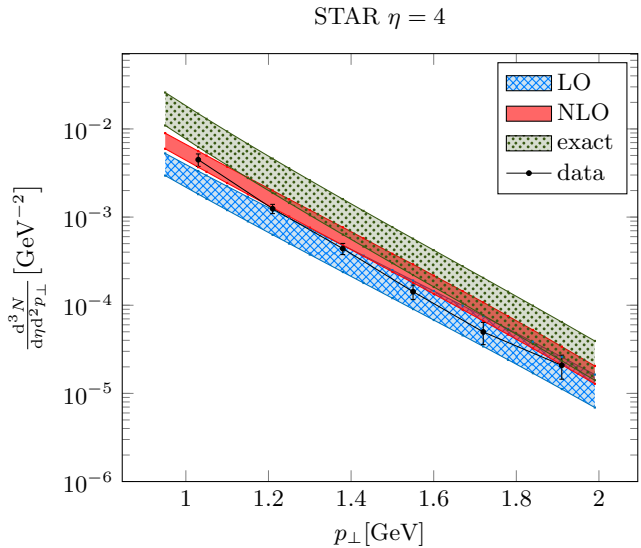

At the end of the day, we can sum up all the contributions in the large $p_\perp$ limit and compare with the BRAHMS data for $y=2.2$ and $y=3.2$ in $\dAu$ collisions at $\sqrt{s}=\SI{200}{GeV}$ as shown in Fig.~\ref{ex}.
In the same figure, we also plot the LO and NLO results computed from the small-$x$ formalism.
We see that the curve with the exact kinematics, which matches the collinear factorisation results, agrees with the data when $p_\perp \gtrsim Q_s$, while it overpredicts the data in the $p_\perp \lesssim Q_s$ region.
This is expected since the perturbative expansion starts to fail in the $p_\perp \lesssim Q_s$ regime, where the saturation formalism takes over and provides decent agreement with the data.
More specifically, we can see that the matching point is around $p_\perp \simeq \SI{1}{GeV}$ for $\eta=2.2$, and it gets up to $p_\perp \simeq \SI{1.5}{GeV}$ for $\eta=3.2$, due to decreasing $x_a$ and increasing saturation momentum $Q_s$.
As shown in Fig.~\ref{ex2}, it is then natural to see that the small-$x$ calculation always gives a good description of the data up to the end of the spectrum for $\eta=4$, while the perturbative result with the exact kinematics overpredicts the data until $p_\perp$ gets to around $\SI{2}{GeV}$.
In addition, we can see that due to the additional positive definite $2\to 2$ contributions from the $q\to q$ and $g\to g$ channels with the exact kinematics, the perturbative curves with the exact kinematics are now always larger than the LO curves.

\section{Summary and Discussions}
In conclusion, we have demonstrated that the saturation formalism expression for forward inclusive hadron production in $\pA$ collisions in the large transverse momentum region can be matched to the corresponding collinear factorization result.
This matching can help to extend the NLO small-$x$ calculation to the large transverse momentum region.
Following this idea, we have proposed the use of the exact kinematics to properly describe available data at large transverse momenta, while the small transverse momentum region can still be accurately described by the full NLO small-$x$ calculation. 

\section*{Acknowledgments}

We thank G. Beuf, G. Chirilli, Y. Kovchegov, A. Mueller, J.W. Qiu and W. Vogelsang for discussions and comments. This work was supported in part by the U.S. Department of Energy under the contracts DE-AC02-05CH11231 and DOE OJI grant No. DE - SC0002145, and by the Polish NCN grant DEC-2011/01/B/ST2/03915.

\appendix

\section{Derivation of Exact Kinematics}\label{sec:kinematicappendix}

This appendix briefly outlines the derivation of the exact kinematic relations~\eqref{exk}. We begin with energy-momentum conservation in light-cone coordinates ($p^{\pm} = E \pm p_z$) for the process shown in Fig.~\ref{fig:qqgdiagram}:
\begin{align}
 x p_p^+ &= k^+ + q^+ &
 x_a p_a^- &= k^- + q^- &
 \vec{k}_{g\perp} &= \vec{k}_\perp + \vec{q}_{\perp} \label{eq:momentumconservation}
\end{align}
Combining the definition of the NLO momentum fraction $\xi$, namely $\xi = \frac{k^+}{x p_p^+}$, with the $+$ component of Eq.~\eqref{eq:momentumconservation} and assuming the final-state quark is massless ($k^2 = 0$) gives the result $x = \frac{p_\perp}{z\xi\sqrt{s}}e^{y}$.

The gluon carries a fraction $1 - \xi$ of the parent quark's momentum, i.e. $1 - \xi = \frac{q^+}{x p_p^+}$. Since this is a final state gluon, we take it to be on shell, $q^+ q^- = q_{\perp}^2$. Combining these last two relations with the $-$ component of Eq.~\eqref{eq:momentumconservation} and the definition of $x$ from the previous     paragraph gives
\begin{equation}
 x_a p_a^- = k_\perp e^{-y} + \frac{\xi q_{\perp}^2}{(1 - \xi)k_\perp}e^{-y}
\end{equation}
and then using the transverse component of Eq.~\eqref{eq:momentumconservation} yields the definition of $x_a$ in Equation~\eqref{exk:xa}.

\end{document}